\documentclass[graybox]{svmult}

\usepackage[utf8]{inputenc}
\usepackage[american]{babel}

\usepackage{mathptmx}       
\usepackage{helvet}         
\usepackage{courier}        
\usepackage{type1cm}        
\usepackage{makeidx}         
\usepackage{graphicx}        
\usepackage{multicol}        
\usepackage[bottom]{footmisc}

\usepackage{amsmath}
\usepackage{amssymb}
\usepackage{url}

\makeindex             

\usepackage{setspace}
\usepackage{amsmath}
\usepackage{amssymb}
\usepackage{algorithm,algorithmicx}
\usepackage[noend]{algpseudocode}
\makeatletter
\def\BState{\State\hskip-\ALG@thistlm}
\makeatother
\usepackage{array}
\newcolumntype{C}[1]{>{\centering\let\newline\\\arraybackslash\hspace{0pt}}m{#1}}

\usepackage{xspace}

\newcommand{\algoName}{SDG\xspace}
\newcommand{\algoNameE}{SEDGE\xspace}

\begin{document}

\title*{A generative model for sparse, evolving digraphs}
\author{Georgios Papoudakis and Philippe Preux and Martin Monperrus}
\institute{Georgios Papoudakis \email{giwrpapoud@gmail.com}\hfill~\linebreak
Philippe Preux \email{firstname.lastname@inria.fr}
\at Université de Lille, CRIStAL \& Inria, Villeneuve d'Ascq, France.
\and Martin Monperrus \at KTH Royal Institute of Technology, Sweden \email{firstname.lastname@csc.kth.se}}
\maketitle

\abstract{
Generating graphs that are similar to real ones is an open problem, while the similarity notion is quite elusive and hard to formalize. In this paper, we focus on sparse digraphs and propose \algoName{}, an algorithm that aims at generating graphs similar to real ones. 
Since real graphs are evolving and this evolution is important to study in order to understand the underlying dynamical system, we tackle the problem of generating series of graphs. We propose \algoNameE{}, an algorithm meant to generate series of graphs similar to a real series. \algoNameE{} is an extension of \algoName{}.
We consider graphs that are representations of software programs and show experimentally that our approach outperforms other existing approaches. Experiments show the performance of both algorithms.
}

\section{Introduction}

We wish to generate artificial graphs that are similar to real ones: by ``real'', we mean a graph that is observed in the real world; as we know, there is ample evidence that graphs coming from the real world are not Gilbert, or Erdös-Rényi graphs, but exhibit more structure. The motivations range from pure intellectual curiosity to, for instance, being able to test ideas on a set of graphs when only one is available (the WWW, a social network), or understanding which are the key properties of a graph. 
This paper is considering directed, un-looped, un-weighted, sparse graphs of moderate sizes (number of nodes ranging from 100 to a couple of thousands of nodes); by sparse, we mean that the number of edges is of the order of the number of vertices, and typically scales like $a N$, with $a$ very small with regards to $N$ (say $a \le 10$ to give an idea of its value). We assume weak connectivity of the graph. As a case study, we experiment with graphs extracted from software programs; beyond a better understanding of software programs, such graphs may be used \textit{e.g.} to improve software development and track the sources of bugs \cite{SQJ2017,SCAM2016}. 

To generate a graph in this context, one may use an algorithm that builds a graph given the degree distribution of the real graph (see \cite{KleitmanWang1973} and followers), or its adjacency matrix \cite{curveball}, or some other structure (see \cite{Staudt2017} and references therein). As we wish to understand and model the creation of the graph, and as real graphs are often dynamic, we are more interested in a second type of algorithms that build a graph incrementally. Another motivation is that we do not want to generate graphs that have the exact same number of vertices, or the exact same degree distribution, or anything identical to the real one. A reason for this is that if we consider the degree distribution, two graphs having the same degree distribution may be very different regarding their other properties; in the other way around, two graphs that have more or less slightly different degree distribution, may have very similar properties. The properties we are interested in are of various natures: connectivity, diameter, average path length, transitivity, modularity, assortativity, spectral properties, degree distribution. Furthermore, when considering degree distribution or spectral properties, it is not clear how to meaningfully measure the difference between two degree distributions: mean squared distance, Kolmogorov-Smirnov statistics, Kullback-Leibler divergence, Jensen-Shannon distance. Finally, an important property of the generator is its stability. We identify two types of stability: the first is that for a given set of parameters, the graphs that are generated should have approximately the same characteristics; the other is that the graphs generated by a set of parameters should not change too much when the value of the parameters change a bit (sort of continuity of the properties of the generated graphs in the space of parameters of the generator).

Modeling and generating static graphs is important, but we are really interested in modeling the evolution of a graph. Though some works exist \cite{Holme2015}, the issues mentioned above take yet another aspect when considering the evolution of a graph. We see the evolution of a graph as time series of graphs, that is a set of couples $\{ (t, g) \}$. We wish to generate the whole series of graphs with a single algorithm. Succeeding in this endeavor, we would access to general properties of the graph and the evolution process, as well as being able to predict the next graphs.

The content of this paper is as follows: in section \ref{sec:algo}, we propose the Sparse Digraph Generator (\algoName{}) which is an algorithm that generates graphs that fit our requirements; we then show that the degree distribution follows a power law distribution; we also show that the in-degree and the out-degree distributions are not identical, something often observed in real digraphs. Then, we put \algoName{} to the test: we introduce the real graphs we work with and show how our generator performs. As we are interested in the modeling of the evolution of a dynamic graph, we introduce Sparse Evolving Digraph GEnerator (\algoNameE{}) which is an incremental version of \algoName{} in section \ref{sec:algoIncremental} and put it to the test in section \ref{sec:expe2}. Then, we conclude and draw some final remarks.

For the sake of reproducible research, all the experiments may be reproduced with the material freely available at \url{https://github.com/papoudakis/sparse-digraph-generator}.

\section{The Sparse Digraph Generator: \algoName}
\label{sec:algo}

We present a novel algorithm that aims at generating sparse digraphs. It is outlined in algorithm \ref{algo:generativeModel}. \algoName{} starts by creating a digraph made of $N$ isolated nodes and then, at each iteration, it adds a link between two nodes.
To add a link, \algoName{} selects two nodes, one as output, and the other as an input node. The selection of either node is performed either at random or following a preferential attachment rule. 
\begin{algorithm}
	\caption{Outline of \algoName{}}
	\begin{algorithmic}[1]
		\State \textit{\textbf{Input:} Number of nodes: N}
		\State \textit{\textbf{Input:} Number of edges: E (assumed to be $\ll N^2$)}
		\State \textit{\textbf{Input:} Parameters $e_1$ and $e_2$, both in the range $[0,1]$}
		\State \textit{\textbf{Output:} Generated graph \textit{G}}
		\State \textit{G} $\gets$ \textit{DiGraph (with N nodes and no edge)}

		\For {t $\in \{ 1, ..., E\}$}
        \State 
        \Comment {Selection of the node that the edge will start from}
		\State \textit{\textbf{With} probability $e_1$:} \textit{out} $\gets$ \textit{select\_a\_node\_uniformly\_at\_random()}
		\State \textit{\textbf{Otherwise}:} \textit{out} $\gets$ \textit{select\_a\_node\_by\_preferential\_attachment}
        
        \State
        \Comment {Selection of the node that the edge will end to}
	\State \textit{\textbf{With} probability $e_2$:} \textit{ in} $\gets$ \textit{select\_a\_node\_of\_in-degree\_0()}
	\State \textit{\textbf{Otherwise}:} \textit{in} $\gets$ \textit{select\_a\_node\_by\_preferential\_attachment}
		\State \textit{G.add\_edge(out, in)}
		\EndFor
		\Return \textit{G}
	\end{algorithmic}
    \label{algo:generativeModel}
\end{algorithm}

We consider sparse digraphs in which the number of edges $E$ is $a N$, where $a \in (1,10)$. Such digraphs are quite common in applications and they are quite specific with regards to their properties: for instance, there is usually a very small number of paths to navigate from one node to another.
It is often the case that the in-degree and the out-degree distributions do not have the same shape. \algoName{} achieves this: is $e_1 \neq e_2$, the parameters of the power law of in-degree and out-degree distributions are different.

The selection of a node to connect to or from is either uniformly at random (among all nodes at line 8, among nodes of in-degree 0 at line 11), or with a probability proportional to the degree of the node, that is we use a linear preferential attachment rule.

In the rest of this section, we derive the form of the in-degree and out-degree distributions resulting from \algoName{}. We show that both distributions follow a power law, though of different parameters.

\subsection{The in-degree distribution}

After the completion of the $t^{\mbox{\small th}}$ iteration of \algoName{}, the graph is made of $t$ edges. So, the probability for a node of degree $k$ to be selected by linear preferential attachment is $\frac{k}{t}$. Additionally, we assume that $e_2 < \frac{N}{E}$, so the expected number of nodes that have in-degree 0 is bigger than 0, $\mathbb{E}[N - e_2E] > 0$


Let $D_k(t)$ be the number of nodes with in-degree $k$ at timestep $t$. For $k>1$, $D_k(t)$ decreases at timestep $t$ only if a node with in-degree $k$ is selected due to preferential attachment (line 12). So the probability that $D_k$ decreases at iteration $t$ is:
\begin{equation}
\underbrace{(1-e_2)}_{\mbox{\scriptsize probability of selecting a node by preferential attachment}}\underbrace{\frac{k}{t}}_{\mbox{\scriptsize probability of choosing a degree k node}}D_k(t)
\end{equation}

Similarly, $D_k(t)$ increases only if a node with in-degree $k-1$ is selected due to preferential attachment. So the probability that $D_k$ increases at iteration $t$ is:
\begin{equation}
(1-e_2)\frac{(k-1)}{t}D_{k-1}(t)
\end{equation}

Let $d_k(t)=\mathbb{E}[D_k(t)]$. It follows that the expected change in the number of nodes of degree $k$ at iteration $t$ is:
\begin{equation}
	d_k(t+1) - d_k(t) = (1-e_2)\frac{(k-1)d_{k-1}(t) - kd_k(t)}{t}
\end{equation}






We set $c_2=1-e_2$ and we assume that $d_k(t) = p_k t$ so we get:
\begin{equation}
p_k = c_2((k-1)p_{k-1} - kp_k)
\end{equation}


\begin{equation}
p_k = \bigg{(}1 - \frac{(1 + c_2)/c_2}{1/c_2 + k}\bigg{)}p_{k-1} 
\end{equation}
Assuming that $ k \gg \frac{1}{c_2}$ and using the binomial approximation we come up with:
\begin{equation}
p_k  	\approx \bigg{(}1 - \frac{(1 + c_2)/c_2}{ k}\bigg{)}p_{k-1} \approx \bigg{(}\frac{k-1}{k}\bigg{)}^ \frac{1 + c_2}{c_2}p_{k-1} 
\end{equation}
Finally, we calculate the values of $p_0$ and $p_1$ and we iterate the equation until $k=2$.  
\begin{equation}
p_k \approx \bigg{(}\frac{k-1}{k}\bigg{)}^ \frac{1 + c_2}{c_2} \bigg{(}\frac{k-2}{k-1}\bigg{)}^ \frac{1 + c_2}{c_2} ... \bigg{(}\frac{1}{2}\bigg{)}^ \frac{1 + c_2}{c_2} p_1 
\end{equation}

\begin{equation}
p_k \approx p_1k^ {-\frac{1 + c_2}{c_2}}
\end{equation}


\subsection{The out-degree distribution}

In this section, $D_k(t)$ is the number of nodes with out-degree $k$ at iteration $t$.
Starting with the same assumptions as before, we can write that the number of nodes with out-degree distribution $k$ decreases if a node with out-degree $k$ is selected due to preferential attachment with probability $1-e_1$ or if such a node is selected from a uniform distribution with probability $e_1$. This second possibility is different from the analysis we did for the in-degree distribution. So, the probability that $D_k$ decreases at iteration $t$ is:
\begin{equation}
e_1\frac{D_k(t)}{n}  + (1-e_1)k\frac{D_k(t)}{t}
\end{equation}

Similarly, $D_k(t)$ increases with probability:
\begin{equation}
e_1\frac{D_{k-1}(t)}{n}  + (1-e_1)(k-1)\frac{D_{k-1}(t)}{t}
\end{equation}

After following the same steps as before we end up with:

\begin{equation}
d_k(t+1) - d_k(t) = (1-e_1)\frac{(k-1)d_{k-1}(t) - kd_k(t)}{t}
+ e_1\frac{d_{k-1}(t) - d_k(t)}{N}
\end{equation}






Assuming that the solution is like $d_k(t) = p_k t$ and by setting $c_1 = 1-e_1$, we can prove that at the final timestep $t=E$:
\begin{equation}
p_k \approx p_1 (k + \frac{(1-c_1)}{c_1} \frac{E}{N}) ^ {-\frac{1 + c_1}{c_1}}
\end{equation}

\subsection{Discussion \& Related Work}

We have shown that the in-degree and the out-degree distributions of the graphs generated by \algoName{} exhibit a power law. This may come as a surprise to the reader, well aware of earlier works, such as \cite{BA-Science}. Indeed, our graph is not growing, keeping a set of $N$ nodes, connecting them along the iterations of the algorithm. However, the departure from a power law is expected when the number of iterations is approximately $N^2$, that is when the graph gets dense. However, as we emphasized it earlier, we only consider sparse graphs, and the number of iterations, hence the number of edges, remains ${\cal O}(N)$, hence much less than $N^2$.

It is worth noting that the power law coefficients of graphs generated by \algoName{} are the same as those of graphs produced by Bollobas \textit{et al.}, though the algorithms are slightly different. Actually Bollobas \textit{et al.} results come as special cases of our analysis.

\algoName{} departs from the usual Barabasi-Albert type of algorithms because it generates directed graphs. Strictly speaking, our algorithm generates a variant of a Price graph \cite{NewmanSurvey} and setting $e_1$ to $0$, $e_2$ to $1$, a kind of Price's algorithm which adds one edge at a time is recovered. \algoName{} comes very close to the one studied by Bollobas \textit{et al.} \cite{bollobas2003} though only \algoName{} is able to add two vertices at once, in a single iteration. 

\section{Experimental study of \algoName}
\label{sec:expe}

In this experimental section, we mainly study two questions:
\begin{itemize}
\item which algorithm performs the best to produce graphs that are similar to some real graphs?
\item the stability of \algoName{} with regards to its parameters.
\end{itemize}

We compare our algorithm with GDGNC \cite{1410.7921} where it is shown to be the best graph generator available in the context of software graphs. We also compare our model with Bollobas \textit{et al.}'s since they are quite similar: it is interesting to check how the small difference in these 2 algorithms convert into difference of performance. We have compared \algoName{} with other algorithms (Kronecker graphs, ...) but since they perform poorly and due to space limitations, we do not report them. 
The experiments are performed with 10 major software programs taken from the maven dataset \cite{mavenMSRpaper}. Table \ref{table:dataset} summarizes the basic features of our dataset.
\begin{table}[H]
	\centering
    \begin{tabular}{|p{3.5cm}|c|c|c|c|}
    \hline
    Software (version) & Nodes & Edges & Edges/Nodes & Diameter \\
    \hline
      	ant (1.5.1)& 266 & 1427 & 5.36 & 6 \\
		findbugs (0.6.4) & 56 & 183 & 3.27 & 5\\
		freemarker (1.5.3) & 76 & 358 & 4.71 & 7\\
		hibernate (1.2) & 365 & 1916 & 5.25 & 7\\
		htmlunit (1.10) & 219 & 934 & 4.26 & 7\\
		jasperreports (3.1.2) & 1139 & 7460 & 6.54 & 7\\
		jparsec (0.2.2) & 75 & 203 & 2.71 & 5\\
		ojb (0.5.200) & 179 & 766 & 4.28 & 6\\
		pmd\_jdk14 (4.1.1) & 521 & 3049 & 5.85 & 8\\
		spring\_core (1.0.1) & 112 & 337 & 3.01 & 7\\
	\hline
	\end{tabular}
	\caption{Statistics of the dataset used in the experiments reported in section \ref{sec:expe}.}
    \label{table:dataset}
\end{table}

In the literature, the measure of similarity between two graphs is not very well defined. In this paper, we measure the similarity between the generated graph ($gg$) and the original graph ($go$) using the following set of metrics:
\begin{itemize}
\item The Kolmogorov-Smirnov statistic (KS) of the in-degree and out-degree distributions.
Let $CDF_g$ denote the cumulative degree distribution function of a graph $g$, so that $CDF_g(k) = \sum_{i\le k} D_k$ where $D_k$ is the degree distribution of graph $g$. Then, $KS=\max_k{|CDF_{gg} (k) - CDF_{go} (k)|}$. We denote $KS_{in}$ (resp.\@ $KS_{out}$) the KS statistics regarding in-degree (resp.\@ out-degree) distribution.

\item The mean squared distance (MSD) of the sorted in-degree and out-degree distributions. For each generated graphs $g$ we consider the in-degree and out-degree of each node, sort these two lists to obtain $d_{in,g}$ and $d_{out,g}$. Then:
$MSD_{in} = \frac{\sum_i{(d_{in,gg}(i)-d_{in,go}(i))^2}}{N} $ and  $MSD_{out} = \frac{\sum_i{(d_{out,gg}(i)-d_{out,go}(i))^2}}{N}$.

The MSD can only be used for \algoName{} and GDGNC because they generate the same number of nodes as the original graph. On the contrary, Bollobas \textit{et al.}' model does not necessarily produce graphs with the same number of nodes.

\end{itemize}

We perform a grid search in order to determine the parameters of each model that best fit for each graph. \algoName{} and GDGNC are optimized to minimize the maximum value between $MSD_{in}$ and the $MSD_{out}$: $minimize \{max(MSD_{in}, MSD_{out}) \}$. As $MSD_{in}$ and $MSD_{out}$ are irrelevant for it, Bollobas \textit{et al.} model is optimized to minimize the KS statistic. This may be seen as a caveat in our experiments, but we provide ample observations below to convince the reader that if we were tuning the parameters of the 3 models with the same metrics, the conclusions of the experiments would not change much.
The experiments presented in table \ref{table:degreeExpe} below are performed with the optimal parameters for each software, averaged over 100 generated graphs. Table \ref{table:degreeExpe} provides the average value of KS and MSD for each model and each software.

\begin{table}[H]
	\centering
	\begin{tabular}{|l|c|c|c|c|c|c|c|c|c|c|}
    	\hline
		 & \multicolumn{3}{c|}{$KS_{in}$} & \multicolumn{3}{c|}{$KS_{out}$} & \multicolumn{2}{c|}{$MSD_{in}$} & \multicolumn{2}{c|}{$MSD_{out}$} \\
		\hline
		Software & \algoName{} & GDGNC & Bollobas & \algoName{} & GDGNC & Bollobas & \algoName{} & GDGNC & \algoName{} & GDGNC \\
		\hline
		ant & 0.26 & \textbf{0.24} & 0.39 & \textbf{0.16} & 0.17 & 0.34 & \textbf{17.4} & 30.45 & \textbf{1.89} & 2.58\\ 
findbugs & \textbf{0.29} & 0.41 & 0.37 & \textbf{0.33} & 0.35 & 0.37 & \textbf{2.32} & 3.74 & \textbf{1.24} & 2.65\\ 
freemarker & \textbf{0.23} & \textbf{0.23} & 0.4 & 0.48 & 0.49 & \textbf{0.38} & \textbf{3.11} & 6.14 & \textbf{4.89} & 6.76\\ 
hibernate & 0.38 & 0.41 & \textbf{0.33} & \textbf{0.22} & 0.32 & 0.32 & \textbf{14.38} & 21.87 & \textbf{3.14} & 9.23\\ 
htmlunit & \textbf{0.37} & \textbf{0.37} & 0.42 & \textbf{0.31} & 0.36 & 0.44 & \textbf{12.67} & 20.68 & \textbf{3.92} & 8.45\\ 
jasperreports & \textbf{0.24} & \textbf{0.24} & 0.28 & 0.35 & 0.43 & \textbf{0.29} & \textbf{32.37} & 97.72 & \textbf{16.1} & 37.43\\ 
jparsec & \textbf{0.22} & \textbf{0.22} & 0.41 & \textbf{0.36} & 0.47 & 0.42 & \textbf{0.69} & 2.72 & \textbf{4.6} & 9.98\\ 
ojb & 0.26 & \textbf{0.25} & 0.44 & \textbf{0.21} & 0.27 & 0.4 & \textbf{3.6} & 6.36 & \textbf{0.77} & 3.41\\ 
pmd\_jdk14 & \textbf{0.27} & 0.28 & 0.28 & 0.5 & 0.56 & \textbf{0.41} & \textbf{14.92} & 114.9 & \textbf{32.08} & 54.54\\ 
spring\_core & \textbf{0.36} & 0.4 & 0.4 & \textbf{0.23} & 0.34 & 0.3 & \textbf{2.54} & 4.67 & \textbf{1.2} & 3.72\\
		\hline
	\end{tabular}
	\caption{Comparison of \algoName{} with GDGNC and Bollobas \textit{et al.} in terms of MSD and KS for 10 Java software graphs. Bold faces indicate best results.}
    \label{table:degreeExpe}
\end{table}


We can clearly see that \algoName{} performs better than both GDGNC and Bollobas \textit{et al.} model. Additionally, \algoName{} is much more stable than the other models. That means that given the parameters of the generator the graphs that are produced are similar. In table \ref{table:meanStd}, we give the average of the standard deviation for the experiments that appear in table \ref{table:degreeExpe}.

\begin{table}[h!]
 \centering
  \begin{tabular}{|c|c|c|c|c|}
    	\hline
        Model & $KS_{in}$ & $KS_{out}$ & $MSD_{in}$ & $MSD_{out}$ \\
        \hline
        \algoName{} & $0.093 \pm 0.012$ & $0.084 \pm 0.023$  & $3.94 \pm 2.55$ & $1.33 \pm 1.07$\\
        GDGNC & $0.091 \pm 0.01$ &  $0.081 \pm 0.013$ & $19.78 \pm 26.6$  & $4.01 \pm 3.89$\\
        Bollobas & $0.102 \pm 0.029$ & $0.099 \pm 0.025$  &  &\\
        \hline
\end{tabular}
\label{table:meanStd}
\caption{Mean and standard deviation of standard deviation values of MSD and KS on 10 Java software graphs.}
\end{table}
From table \ref{table:meanStd} we can see the standard deviation values of \algoName{} are on the same level or smaller than both GDGNC and Bollobas \textit{et al.} But the most important property of \algoName{} is that it can create graphs similar to the original one without the parameter optimization process, that both other models require in order to perform decently. For each software, we generate 100 graphs and we compute the average $KS_{in}$, $KS_{out}$, $MSD_{in}$, and $MSD_{out}$. All the experiments are performed with the same values $e_1=0.45$ and $e_2=\frac{N}{E} - 0.05$ for all software graphs; these values result from our experiments. Table \ref{table:noTuning} provides the results; in ()'s, we report the ratio between the \algoName{} without and with tuning: \textit{e.g.}, $0.14(0.9)$ is the first row of column $KS_{out}$ means that $KS_{out}$ is $0.14$ without tuning, and $0.14/0.9$ with tuning. The value of $KS$ without tuning may be smaller than with tuning because the parameter tuning is performed to minimize $MSD$.


\begin{table}[h!]
  \centering
  \begin{tabular}{|l|C{20mm}|C{20mm}|C{20mm}|C{20mm}|}
    	\hline
        Software & $KS_{in}$ & $KS_{out}$ & $MSD_{in}$ & $MSD_{out}$ \\
        \hline
ant & 0.25 (1.0) & 0.14 (0.9) & 20.54 (1.2) & 0.89 (0.5) \\ 
findbugs  & 0.3 (1.0) & 0.34 (1.0) & 2.66 (1.1)  & 1.34 (1.1)\\ 
freemarker & 0.24 (1.0)& 0.46 (1.0) & 3.43 (1.1) & 5.31 (1.1)\\ 
hibernate  & 0.29 (0.8) & 0.3 (1.4) & 27.16 (1.9) & 13.27 (4.2)\\ 
htmlunit  & 0.33 (0.9) & 0.29 (0.9) & 12.84 (1.0) & 5.24 (1.3)\\ 
jasperreports & 0.21 (0.9) & 0.43 (1.2)  & 119.42 (3.6) & 49.16 (3)\\ 
jparsec  & 0.25 (1.1) & 0.42 (1.2) & 1.52 (2.2) & 8.41 (1.8)\\ 
ojb  & 0.33 (1.3) & 0.27 (1.3) & 13.47 (3.7) & 2.36 (3.1)\\ 
pmd\_jdk14  & 0.33 (1.2) & 0.54 (1.1) & 61.67 (4.1) & 45.43 (1.4)\\ 
spring\_core  & 0.3 (0.8) & 0.25 (1.1) & 2.63 (1.0) & 2.43 (2.0) \\
		\hline
  \end{tabular}
  \caption{MSD and KS without tuning parameters: numbers in ()'s gives the ratio between the measurement without tuning and the measurement with tuning.}
  \label{table:noTuning}
\end{table}
From table \ref{table:noTuning} we see that in most cases, \algoName{}, without parameter tuning, performs better than both GDGNC and Bollobas \textit{et al.} model after parameter tuning.
Another very nice property is that the performance does not change very much as the value of a parameter is changing: there is some sort of continuity of the performance of \algoName{} with regards to the value of parameters. This is a very nice property, as this implies that to tune the parameters of \algoName{}, a coarse grid search is enough and computationally cheaper.

\begin{figure}[!htb]
\minipage{0.32\textwidth}
  \includegraphics[width=\linewidth]{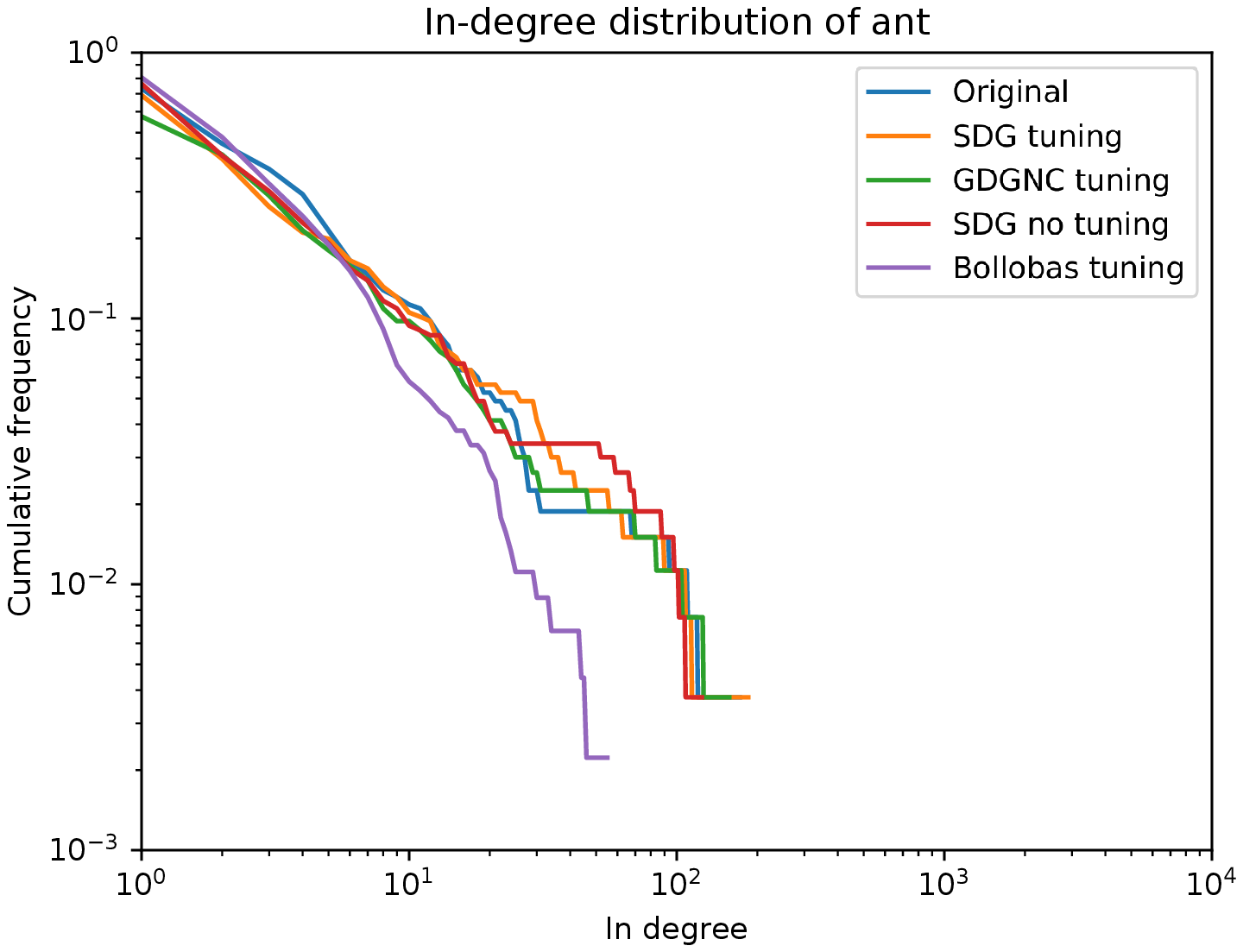}
\endminipage\hfill
\minipage{0.32\textwidth}
  \includegraphics[width=\linewidth]{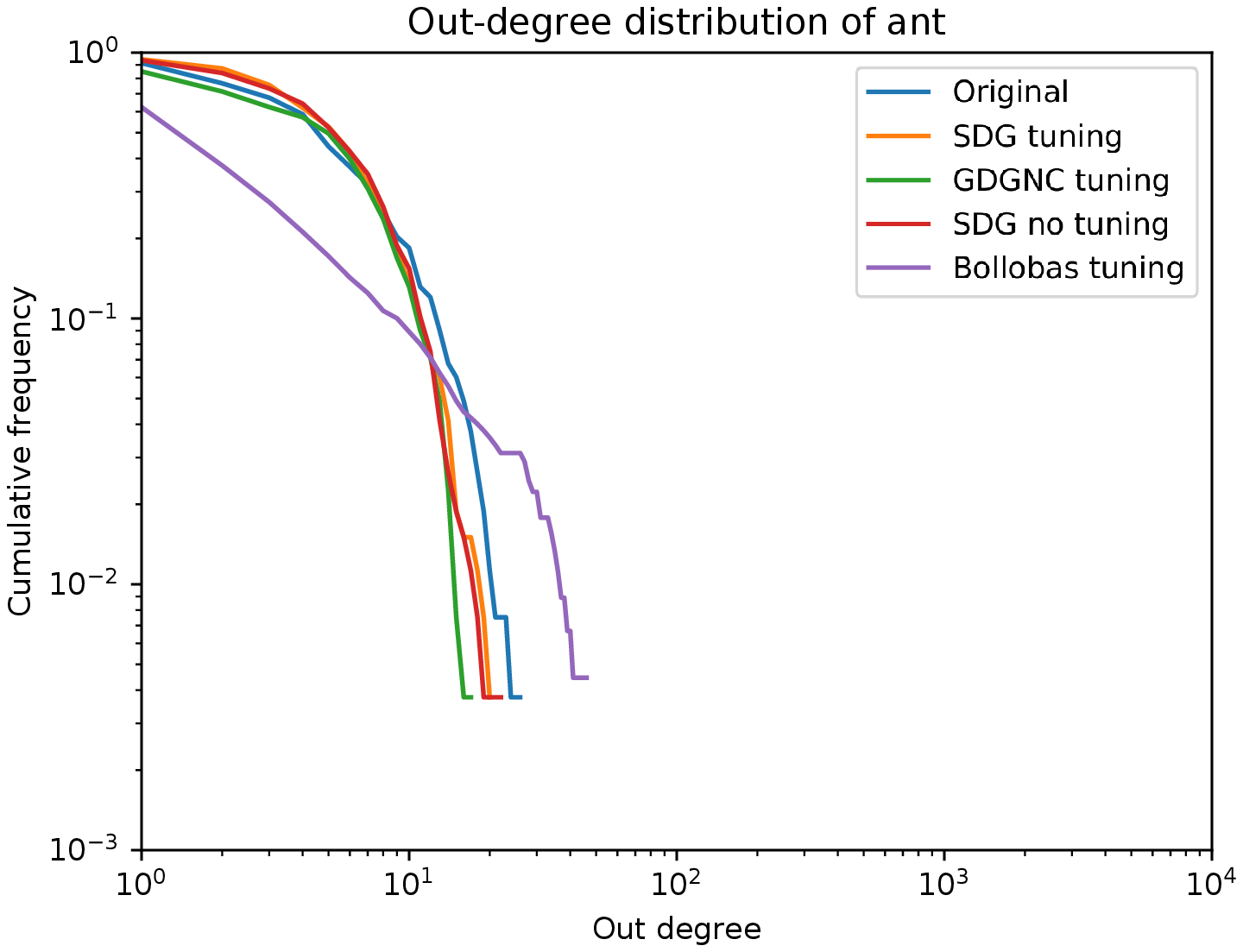}
\endminipage\hfill
\centering
\minipage{0.32\textwidth}%
  \includegraphics[width=\linewidth]{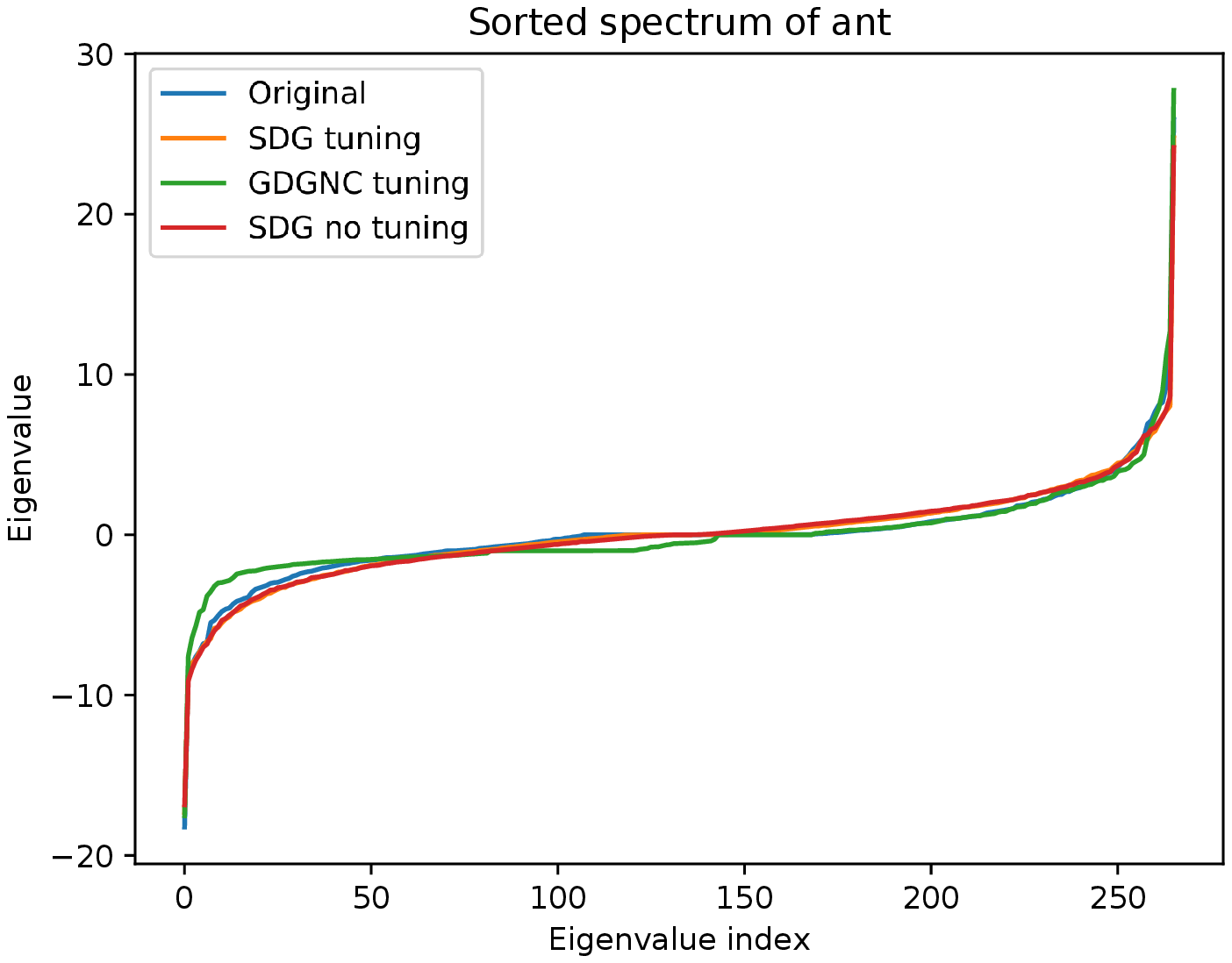}
\endminipage
\caption{In-degree distribution, out-degree distribution, and spectrum for the real graph and the generated graphs.}
\label{fig:in_out_spec}
\end{figure}

Figure \ref{fig:in_out_spec} provides a graphical illustration of these measurements: we plot the in-degree distribution, the out-degree distribution, and the spectra of the adjacency matrix for the real graph and for the graphs generated by each algorithm we compare to.

To conclude this part, let us stress that \algoName{} uses two pieces of information: the number of nodes $N$ and the number of edges $E$. We have shown that \algoName{} produces graphs which degree distributions follow power laws. When we want to generate graphs similar to a real one, both $N$ and $E$ are available, and we have shown that $e_1$ and $e_2$, the parameters of \algoName{}, are not that important to obtain satisfying graphs. Another point is that the occasional addition of 2 nodes instead of 1 seems beneficial since this is the only difference between \algoName{} and Bollobas \textit{et al.} approach.

Finally, it is important to refer to the metrics we use to compare graphs and the metrics we use to optimize the parameters of the algorithm. As said earlier, it is not known how to assess the similarity of two graphs using a single metric; instead, we use a series of metrics (and more may be used) to formalize the idea of similarity between two graphs. The metrics we use are recognized as very important to characterize a graph: the degree distribution, and the spectrum. We have found that optimizing using the degree distributions leads to better results. We see that as a primary observation, other spectral information might be used, and other properties may be used too. Furthermore, a combination of metrics may be optimized or used to judge the similarity: this is left as future work.

\section{\algoNameE: modeling the evolution of a graph}
\label{sec:algoIncremental}

We consider a model of evolution of the real graph that is version-oriented. As the real graphs we consider are software, considering a sequence of versions of a software, the graphs along this sequence evolve by part: by that, we mean that the set of nodes and the set of edges evolve by chunks: from one graph to the next one (one version of a software to the next one), a set of nodes are added, some nodes are removed, and it is the same for the edges. So, we consider an algorithm that takes a graph as input, and then adds a set of nodes and a set of edges, possibly removing some existing nodes and edges. 

We propose the ``Sparse Evolving Digraph GEnerator'' \algoNameE{} (see algorithm \ref{algo:generativeModelEvol}), a model to capture the evolution of software graphs based on the generative model that we proposed in section \ref{sec:algo}. \algoNameE{} is an extension of \algoName. It distinguishes existing nodes from new nodes. At each timestep, \algoNameE{} chooses two nodes to connect, sampling them from either set of nodes, based on 2 parameters that act as probabilities $\alpha$ and $\beta$. 

\begin{algorithm}
	\caption{\algoNameE: a generative model for sparse digraph evolution. The \textsc{sample\_a\_node} samples nodes in exactly the same way algorithm \ref{algo:generativeModel} does. new\_nodes refers to the N\_new nodes that are added to the current graph. all\_nodes refers to all nodes of the new graph.}
	\begin{algorithmic}[1]
		\State \textit{\textbf{Input:} Number of nodes to add N\_new}
		\State \textit{\textbf{Input:} Number of edges to add E\_new}
		\State \textit{\textbf{Input:} Parameters $\alpha, \beta, e_1, e_2$, all in the range $[0,1]$}
        \State \textit{\textbf{Input:} Current graph G\_cur}
		\State \textit{\textbf{Output:} Generated graph \textit{G\_new}}
        
        \Function{sample\_a\_node} {so, si, $e_1$, $e_2$}
          \State \textit{\textbf{With} probability $e_1$:} \textit{out} $\gets$ \textit{select\_a\_node\_uniformly\_at\_random(so)}
          \State \textit{\textbf{Otherwise}:} \textit{out} $\gets$ \textit{select\_a\_node\_by\_preferential\_attachment (so)}
	      \State \textit{\textbf{With} probability $e_2$:} \textit{ in} $\gets$ \textit{select\_a\_node\_of\_in-degree\_0(si)}
	      \State \textit{\textbf{Otherwise}:} \textit{in} $\gets$ \textit{select\_a\_node\_by\_preferential\_attachment (si)}
          \State \textbf{return} (in, out)
        \EndFunction
        \State \textbf{End function} 
		\State \textit{G\_new} $\gets$ \textit{G\_cur.add\_nodes(N\_new)}

		\For {t $\in \{ 1, ..., E\_{new} \} $}
		\State \textit{\textbf{With} probability $\alpha$:}
          (in, out) $\gets$ \Call{sample\_a\_node}{all\_nodes, new\_nodes, $e_1$, $e_2$}
		\State \textit{\textbf{With} probability $\beta$:}
          (in, out) $\gets$ \Call{sample\_a\_node}{new\_nodes, all\_nodes, $e_1$, $e_2$}
		\State \textit{\textbf{Otherwise}:}
          (in, out) $\gets$ \Call{sample\_a\_node}{all\_nodes, all\_nodes, $e_1$, $e_2$}
		\State \textit{G\_new.add\_edge(out, in)}
		\EndFor
		\Return \textit{G\_new}
	\end{algorithmic}
    \label{algo:generativeModelEvol}
\end{algorithm}

\section{Experimental study of \algoNameE}
\label{sec:expe2}

In this section, we evaluate the ability of \algoNameE{} to capture the software evolution. For the experiments, we use 10 pairs of consecutive versions of software graphs\footnote{(ant.1.4.1$\rightarrow$ant.1.5), (commons\_collections.20030418.083655$\rightarrow$commons\_collections.20031027. 000000), (hibernate.2.0.3$\rightarrow$hibernate.2.1.1), (jasperreports.0.6.7$\rightarrow$jasperreports.1.0.0), (jasperre- ports.1.0.3$\rightarrow$jasperreports.1.1.0), (ojb.0.8.375$\rightarrow$ojb.0.9), (ojb.0.9.5$\rightarrow$ojb.0.9.6), (spring.1.0 $\rightarrow$spring.1.1), (wicket.1.0.3$\rightarrow$wicket.1.1), (wicket.1.1.1 $\rightarrow$ wicket.1.2)} from the maven dataset. With the term ``first graph'', we refer to the first version of the software and with the term ``second graph'' to the second version. In each pair of these graphs, the second graph has at least $20\%$ more nodes than the first graph. 

The degree distributions and the spectrum of the graphs of two successive versions are close. For this reason, in order to perform a better evaluation of \algoNameE{} we compute KS and MSD only for the new nodes: doing so, we amplify the difference between the two versions. In table \ref{table:evolTuning}, we report on the values of KS and MSD averaged over 100 experiments, for each real graph, given the optimal parameters of the model.

\begin{table}[h!]
  \setlength{\tabcolsep}{0pt}
  \centering
  \begin{tabular}{|l|C{8mm}|C{9mm}|C{17mm}|C{17mm}|C{17mm}|C{17mm}|}
    	\hline
       First Software & $N_{new}$ & $E_{new}$ & $KS_{in}$ & $KS_{out}$ & $MSD_{in}$ & $MSD_{out}$ \\
        \hline
ant.1.4.1  & 116 & 665 & 0.29 (0.8)  & 0.4 (1.1)  & 5.57 (2.9)  & 2.37 (0.6) \\ 
commons.20030418  & 118 & 385 & 0.41 (1.1)  & 0.39 (0.9)  & 0.99 (1.5)  & 1.05 (0.9) \\ 
hibernate.2.0.3  & 92 & 853 & 0.39 (0.6)  & 0.29 (1.0)  & 3.52 (12.4)  & 3.22 (1.0) \\ 
jasperreports.0.6.7 & 170 & 1100 & 0.23 (1.1)  & 0.2 (1.2)  & 15.39 (1.2)  & 6.08 (1.8) \\ 
jasperreports.1.0.3  & 117 & 1214 & 0.19 (0.9)  & 0.25 (1.0)  & 22.1 (2.7)  & 9.2 (0.9) \\ 
ojb.0.8.375 & 100 & 555 & 0.31 (0.9)  & 0.39 (1.0)  & 5.72 (1.6)  & 1.27 (1.0) \\ 
ojb.0.9.5  & 120 & 586 & 0.47 (1.0)  & 0.36 (1.0)  & 1.51 (0.6)  & 2.4 (3.5) \\ 
spring.1.0  & 199 & 830 & 0.36 (1.0)  & 0.4 (0.8)  & 2.66 (3.7)  & 1.17 (3.2) \\ 
wicket.1.0.3	  & 96 & 569 & 0.36 (0.9)  & 0.32 (1.0)  & 1.21 (26.4)  & 1.93 (1.4) \\ 
wicket.1.1.1  & 235 & 1800 & 0.25 (1.0)  & 0.2 (1.1)  & 4.92 (1.0)  & 2.75 (2.8) \\ 
		\hline
  \end{tabular}
  \caption{MSD and KS for 10 evolutions of software graphs of \algoNameE{}, averaged over 100 runs for each software. We also run the same experiments without tuning parameters: numbers in ()'s gives the ratio between the measurement without tuning and the measurement with tuning: a value below 1 means that it is better without tuning, above 1 that it is worse.}
  \label{table:evolTuning}
  \label{table:evol}
\end{table}

\algoNameE{} has the same fundamental property \algoName{} has: it can capture the structure of the evolved network without tuning its parameters. As in table \ref{table:noTuning}, the values in ()'s in table \ref{table:evol} gives the ratio between tuning and no tuning. We use $\alpha=0.5$, $\beta=0.4$, $e_1=0.45$ and $e_2 = \frac{N}{E} - 0.05$ in the non tuned parameters experiment.


\section{Conclusion and future work}

In this paper, we consider the problem of generating graphs that are similar to real, sparse digraphs. We propose \algoName{} which generates such graphs, exhibiting power law in their degree distributions. We show that \algoName{} performs very well experimentally; furthermore, \algoName{} is stable in terms of parameter tuning: we show that it behaves very well even if we do not perform parameter tuning. Then, we propose an extension named \algoNameE{} which aims at generating series of sparse digraphs that is similar to a series of real graphs. The similarity between two graphs is not well defined; we have used different ways to measure it and we have discussed the influence on the final result of the generator. Other metrics can also be used and will be investigated in the future. We have used \algoName{} and \algoNameE{} with a type of graphs in mind; we have not defined these algorithms using any knowledge on the graphs being modeled: we have designed the algorithms, tested them on some real graphs, and observed the results. We think they may be used for many types of real graphs.
More importantly, considering series of graphs is a very important aspect of our work. As real graphs are evolving, we think that we have to use dynamic models to deal with them to really capture something about the evolution of the real graph, and the understanding of the process underneath.

\subsection*{Acknowledgements}

This work was partially supported by CPER Nord-Pas de Calais/FEDER DATA Advanced data science and technologies 2015-2020, and the French Ministry of Higher Education and Research. We also wish to acknowledge the continual support of Inria, and the stimulating environment provided by the SequeL Inria project-team.

\bibliographystyle{spmpsci}
\bibliography{bibliog}

\end{document}